\documentclass{article}

\usepackage{epsfig} 
\usepackage{amsmath}
\usepackage{amsfonts}
\usepackage{graphicx}

 \usepackage{amssymb}

\date{\today}

\newcommand{\insertplot}[5]{\begin{figure}
 \hfill\hbox to 0.05in{\vbox to #5in{\vfill
 \inputplot{#1}{#4}{#5}}\hfill}
 \hfill\vspace{-.1in}
 \caption{#2}\label{#3}
 \end{figure}}
 \newcommand{\inputplot}[3]{
 \special{ps: plotfile #1}
}
\newcounter{fig}

\newcommand{\ee}{\end{equation}}
\newcommand{\eea}{\end{eqnarray}}
\newcommand{\be}{\begin{equation}}
\newcommand{\bea}{\begin{eqnarray}}

\begin{document}

\title{ 
Electrically charged black branes  in
\\
 ${\cal{N}}=4^+$, $D=5$ gauged supergravity 
} 
  
\author{{\large Yves Brihaye}$^{1}$, 
{\large Ruben Manvelyan}$^{2}$, {\large Eugen Radu}$^{3}$ 
and {\large D. H. Tchrakian}$^{4,~5}$ \\ \\
$^{1}${\small Physique-Math\'ematique, Universite de
Mons-Hainaut, Mons, Belgium}\\
$^{2}${\small Yerevan Physics Institute, Alikhanian Br. St. 2, 0036 Yerevan, Armenia}\\
$^{3}${\small Institut f\"ur Physik, Universit\"at Oldenburg, Postfach 2503
D-26111 Oldenburg, Germany}
  \\
$^{4}${\small School of Theoretical Physics -- DIAS, 10 Burlington
Road, Dublin 4, Ireland} \\
$^{5}${\small  Department of Computer Science,
National University of Ireland Maynooth,
Maynooth,
Ireland}
  }

\maketitle

\begin{abstract} 

We analyze the  properties of  asymptotically AdS electrically charged black brane solutions
in a consistent truncation of the ${\cal{N}}=4^+$, $D=5$ Romans' gauged supergravity which contains 
 gravity, $SU(2)$ and $U(1)$ gauge fields, and a dilaton possessing  a nontrivial
potential approaching a constant negative
value at infinity.
We find that the $U(1)\times U(1)$ solutions become unstable to forming non-Abelian hair. 
These configurations emerge as zero modes of the Abelian solutions
at critical temperature and a critical (nonvanishing) ratio of the electric charges 
and can be viewed as holographic $p-$wave superfluids. 

\end{abstract}

%%%%%%%%%%%%%%%%%%%%%%%%%%%%%%%%%%%%%%%%%%%%%%%%%%%%%%%%%%%%%%%%%%%%%%%%%%%%%%%%%%%%%%%%%
\section{Introduction}
%%%%%%%%%%%%%%%%%%%%%%%%%%%%%%%%%%%%%%%%%%%%%%%%%%%%%%%%%%%%%%%%%%%%%%%%%%%%%%%%%%%%%%%%% 
 
 Recently, considerable effort
has been put into extending AdS/CFT correspondence beyond high-energy physics by constructing
gravity models that are conjectured to be dual to various condensed matter systems.
This has lead to the discovery of holographic superconductors and holographic superfluids,
describing condensed phases of strongly coupled, planar, gauge theories.
Studying such models involves the construction of electrically charged
black holes in an asymptotically AdS spacetime, which, below a critical
temperature become unstable to 
forming hair.
That is, a phase transition occurs to a  superconductor/superfluid
state,
in which a sufficiently large $U(1)$ charge density triggers the spontaneously
breaking of the $U(1)$ symmetry.
Then an operator charged under the $U(1)$ acquires a nonzero expectation value 
(see $e.g.$ \cite{Horowitz:2010gk} for a review of these aspects).

For $p-$wave superconducting black holes, 
the condensing operator is a vector and hence rotational symmetry is broken.
Such black hole solutions have been constructed using either 
charged non-Abelian vector fields \cite{Gubser:2008zu}
or, alternatively, charged two-forms \cite{Aprile:2010ge}.
However, most of the  studies in the literature have assumed an $ad~hoc$ construction of
the lagrangean of the gravitational system, 
without a clear connection with a given supergravity model, 
which makes it rather difficult to describe precisely the application of the AdS/CFT dictionary.
 
At the same time, the gauged supergravity models
generically contain non-Abelian vector fields, which may suggest the
existence of $p-$wave superconducting black hole solutions.
The case of ${\cal{N}}=8$, $D=5$ gauged supergravity \cite{Gunaydin:1985cu}, \cite{Cvetic:2000nc}
is of particular interest, given its connection with ${\cal{N}}=4$ 
$U(N)$ super-Yang-Mills theory in $3+1$ dimensions.
The bosonic sector of this theory consists of the metric, twenty scalars and fifteen $SO(6)$
Yang-Mills (YM) gauge fields\footnote{Note that the field content of the full ${\cal N}=8,~D=5$ gauged supergravity
is richer. However, a number of bosonic fields 
can be consistently  set to zero \cite{Cvetic:2000nc}.}.
Solutions of ${\cal{N}}=8,$ $D=5$  model have been considered by several authors for various consistent truncations,
with subgroups of $SO(6)$ (see $e.g.$ \cite{Cvetic:2009id} and the references therein).
However, 
to our knowledge, to date no attempt has been made to 
construct non-Abelian superconducting black hole solutions in this context.

This paper is aimed as a first step in this direction,
by taking a consistent truncation of the  ${\cal N}=8$ model
corresponding to  ${\cal{N}}=4^+$, $SU(2)\times U(1)$ Romans' gauged supergravity, with a single scalar
field $\phi$ possessing a
 potential $V(\phi)$ which is the sum of two Liouville terms. 
The scalar $\phi$ approaches asymptotically
a constant value $\phi_0$ corresponding to an extremum of the
potential,
$ dV/d\phi \big|_{\phi_0}=0$, which yields an effective cosmological
constant  $\Lambda_{eff}=2V(\phi_0)<0$.
%Then
It turns out that the basic properties of the ${\cal{N}}=4^+$ solutions with non-Abelian fields are
rather similar to those found for pure $D=5$ Einstein-YM-$\Lambda$ system \cite{Manvelyan:2008sv}, \cite{Ammon:2009xh}. 
In particular, we find evidence for the existence,  at low temperatures, of a  superfluid state
with a normalisable
non-Abelian condensate. 

Since Romans' theory arises as a consistent Kaluza-Klein truncation of the type IIB
supergravity on an $S^5$ \cite{Lu:1999bw} and as a consistent compactification of $D=11$
supergravity \cite{Gauntlett:2007sm},
this shows the existence of holographic superfluids
in $D=10,11$ supergravities.

%%%%%%%%%%%%%%%%%%%%%%%%%%%%%%%%%%%%%%%%%%%%%%%%%%%%%%%%%%%%%%%%%%%%%%%%%%%%%%%%%%%%%%%%%
\section{The ${\cal{N}}=4^+$, $D=5$  Romans' gauged supergravity}
%%%%%%%%%%%%%%%%%%%%%%%%%%%%%%%%%%%%%%%%%%%%%%%%%%%%%%%%%%%%%%%%%%%%%%%%%%%%%%%%%%%%%%%%% 
%%%%%%%%%%%%%%%%%%%%%%%%%%%%%%%%%%%%%%%%%%%%%%%%%%%%%%%%%%%%%%%%%%%%%%%%%%%%%%%%%%%%%%%%%
%\subsection{The  model}
%%%%%%%%%%%%%%%%%%%%%%%%%%%%%%%%%%%%%%%%%%%%%%%%%%%%%%%%%%%%%%%%%%%%%%%%%%%%%%%%%%%%%%%%% 
The bosonic sector of the ${\cal{N}}=4$, $D=5$ 
Romans' gauged supergravity  \cite{Romans:1985ps}
consists of gravity, a scalar $\phi$, an
$SU(2)$ YM  potential $A_{\mu}^{(I)}$ (with field strength $F_{\mu \nu}^{(I)}=
\partial_\mu A_\nu^{(I)}-\partial_\nu A_\mu^{(I)}+g_{YM} \epsilon^{IJK}A_\mu^{(J)} A_\nu^{(K)}$ 
and $g_{YM}$ the SU(2) gauge coupling constant), an Abelian potential
$B_{\mu}$ ($f_{\mu \nu}=\partial_\mu B_\nu -\partial_\nu B_\mu$ being the corresponding field strength),
and a pair of two-form fields. These two form fields can
consistently be set to zero, which yields the bosonic part of the action
\begin{eqnarray}
\label{action5} 
I_{bulk}=\frac{1}{4 \pi}\int_{ \mathcal{M}} d^5x   \sqrt{-g}  \Big
(\frac{1}{4} R
-\frac{1}{2}\partial_\mu\phi \,\partial^\mu\phi -\frac{1}{4}{\rm e}^{2a\phi}
F^{(I)}_{\mu\nu} F^{(I) \mu\nu}
-\frac{1}{4} {\rm e}^{-4a\phi} f_{\mu\nu} f^{
\mu\nu}
\\
\nonumber
- \frac{1}{4\sqrt{-g}}\epsilon^{\mu \nu \rho\sigma\tau}
F_{\mu \nu}^{(I)} F_{\sigma \tau}^{(I)} B_{\tau}-V(\phi)\Big) ,
\end{eqnarray}
where $a=\sqrt{\frac{2}{3}}$. Here 
 %\begin{eqnarray}
 %\label{dil-pot} 
$
 V(\phi)= -\frac{1}{8} g_{YM}^2\left( {\rm e}^{-2a\phi}
+2\sqrt{2}\frac{g_{M}}{g_{YM}} {\rm e}^{a\phi} \right)  
 %\end{eqnarray}
$
is the dilaton potential, $g_M$ being the $U(1)$ gauge coupling constant.  

As discussed in \cite{Romans:1985ps}, this theory has three canonical forms, corresponding to
different choices of the gauge coupling constant $g_M$. The case of interest here corresponds to the  ${\cal{N}}=4^+$ version,
in which $g_{M}=g_{YM}/\sqrt{2}$ and thus the dilaton potential is
\begin{eqnarray}
\label{dil-pot-fin} 
V(\phi)= -\frac{1}{8} g_{YM}^2 \left( {\rm e}^{-2a\phi}+2 {\rm e}^{a\phi} \right) .
\end{eqnarray}

The field equations are obtained by varying the action (\ref{action5})  with respect
to the field variables $g_{\mu \nu},A_{\mu}^{(I)},~B_{\mu}$ and $\phi$
\begin{eqnarray}
\label{Einstein-eqs} 
&&
R_{\mu \nu}-\frac{1}{2}g_{\mu\nu}R = 2~ T_{\mu \nu},
~~\nabla^2 \phi-\frac{a}{2}e^{2a\phi}F_{\mu \nu}^{(I)} F^{(I)\mu \nu} 
+ae^{-4a \phi} f_{\mu \nu}f^{\mu \nu}
-\frac{\partial V}{\partial \phi}=0,
\\
\label{YM-eqs}
\nonumber
&&
\partial_{\nu}(e^{-4a \phi}f^{\mu \nu})
-\frac{1}{4 \sqrt{-g}}\epsilon^{\mu \nu \rho \sigma \tau}F_{\nu\rho}^{(I)}F_{\sigma \tau}^{(I)}=0,
~~
D_{\nu}(e^{2a \phi}F^{(I)\mu \nu})-\frac{1}{2 \sqrt{-g}}\epsilon^{\mu \nu \rho \sigma \tau}F_{\nu\rho}^{(I)} f_{\sigma \tau} = 0,
\end{eqnarray}
where the energy-momentum tensor is defined by
\begin{eqnarray}
\label{Tij}
T_{\mu \nu}&=&
\partial_{\mu} \phi \partial_{\nu} \phi
-\frac{1}{2}g_{\mu \nu} \partial_{\sigma}\phi \partial^{\sigma}\phi-
g_{\mu \nu}V(\phi)\ 
\\
\nonumber
&&+
 e^{2a \phi} 
     ( F_{\mu \rho}^{(I)} F_{  \nu \sigma}^{(I)} g^{\rho \sigma}
   -\frac{1}{4} g_{\mu \nu} F_{\rho \sigma}^{(I)} F^{(I)\rho \sigma})
+
 e^{-4a \phi} 
     ( f_{\mu \rho}  f_{  \nu \sigma} g^{\rho \sigma}
   -\frac{1}{4} g_{\mu \nu} f_{\rho \sigma}  f^{ \rho \sigma})
 .
\end{eqnarray}

The scalar potential has exactly one extremum at $\phi=0$, corresponding to an the effective cosmological constant 
%\begin{eqnarray}
%\label{const}
$
\Lambda_{eff}=-\frac{6}{\ell^2}=2 V(0)=-\frac{3}{4}g_{YM}^2.
$
%\end{eqnarray}
Then the effective AdS length scale is fixed by the non-Abelian gauge coupling constant, $\ell= {2\sqrt{2}}/{g_{YM}}$. 

As usual, one supplements (\ref{action5}) with a boundary term
\begin{eqnarray}
\label{bound}
I_{bound}=-\frac{1}{8\pi }\int_{\partial\mathcal{M}} d^{4}x\sqrt{-h}
 K-\frac{1}{8\pi }\int_{\partial\mathcal{M}} d^{4}x\sqrt{-h}
\left(\frac{1}{\ell}W(\phi)+\frac{\ell}{4}\mathcal{R} \right),
%=-\frac{3}{4}.
\end{eqnarray}
where apart from the Hawking-Gibbons surface term we include also a counterterm part which is required to regularize
the total action and the global charges. 
In the above relation,  $\mathcal{R}$ is the Ricci scalar for the induced metric $h$ of the boundary, $K$ is the trace 
of the extrinsic curvature, while  
$W(\phi)={\rm e}^{ 2a\phi}+2 {\rm e}^{-a\phi}$  
(this expression of the counterterm was derived in \cite{Batrachenko:2004fd}, in a more general context).
%is the superpotential for $V(\phi)$,

Then, as in the well known pure-AdS case \cite{Balasubramanian:1999re}, one can
construct a divergence-free boundary stress tensor $\mathsf{T}_{ij}$ from the total action
$I{=}I_{\rm bulk}{+}I_{\rm bound} $ by defining
\begin{eqnarray}
\label{s1}
\mathsf{T}_{ij}=\frac{2}{\sqrt{-h}} \frac{\delta I}{ \delta h^{ij}}
=\frac{1}{8\pi }(K_{ij}-Kh_{ij}-\frac{1}{\ell}h_{ij}W(\phi)+\frac{\ell}{2} E_{ij}),
\end{eqnarray}
where $E_{ij}$ is the Einstein tensor of the  boundary metric, $K_{ij}=-1/2 (\nabla_i n_j+\nabla_j n_i)$ is the extrinsic curvature,
with $n^i$ being an outward pointing normal vector to the boundary.

Thus, a conserved charge 
\begin{equation}
{\mathfrak Q}_{\xi }=\oint_{\Sigma }d^{3}S^{a}~\xi ^{b}\mathsf{T}_{ab},
\label{Mcons}
\end{equation}%
can be associated with a surface $\Sigma$ (with normal 
$n^{a}$), provided the boundary geometry has an isometry generated by a Killing vector $\xi ^{a}$. 
For example, if $\xi =\partial /\partial t$ is a timelike Killing vector, 
then $\mathfrak{Q}_{\xi }$ is the conserved mass ${\cal M}$.  

%%%%%%%%%%%%%%%%%%%%%%%%%%%%%%%%%%%%%%%%%%%%%%%%%%%%%%%%%%%%%%%%%%%%%%%%%%%%%%%%%%%%%%%%%
 \section{The uncondensed phase}
%%%%%%%%%%%%%%%%%%%%%%%%%%%%%%%%%%%%%%%%%%%%%%%%%%%%%%%%%%%%%%%%%%%%%%%%%%%%%%%%%%%%%%%%% 
%%%%%%%%%%%%%%%%%%%%%%%%%%%%%%%%%%%%%%%%%%%%%%%%%%%%%%%%%%%%%%%%%%%%%%%%%%%%%%%%%%%%%%%%%
 \subsection{The solutions}
%%%%%%%%%%%%%%%%%%%%%%%%%%%%%%%%%%%%%%%%%%%%%%%%%%%%%%%%%%%%%%%%%%%%%%%%%%%%%%%%%%%%%%%%% 

We start with a discussion of the basic properties of the 
Abelian black brane solutions of the ${\cal{N}}=4^+$
Romans' model.
They can be found as a particular limit of the black holes obtained in
\cite{Behrndt:1998jd} in the so-called STU model. 
In the general case these black holes 
possess three different $U(1)$ charges and two independent scalars.
After setting one scalar to zero and taking two gauge fields to be equal, one finds  after a suitable
field redefinition, the following black brane solution of the eqs. (\ref{Einstein-eqs})-(\ref{Tij}): 
\begin{eqnarray}
\label{metricU1xU1-1} 
ds^2={\cal H}(r)^{1/3}\left(\frac{dr^2}{f(r)}+r^2 (dx^2+dy^2+dz^2) \right)-{\cal H}(r)^{-2/3}f(r) dt^2,
\end{eqnarray}
with 
\begin{eqnarray}
\label{metricU1xU1-2} 
 {\cal H}(r)=H^2(r) K(r),~~H(r)=1+\frac{2Q^2}{M r^2},~~ 
 K(r)=1+\frac{4q^2}{M r^2},~~ f(r)=-\frac{M}{r^2}+\frac{1}{8}g^2_{YM}r^2 {\cal H}(r)
\end{eqnarray}
%\begin{eqnarray}
%\label{metricU1xU1-3} 
%
%\end{eqnarray}
and the matter fields 
\begin{eqnarray}
\label{metricU1xU1-5} 
&&
\phi(r)=\frac{1}{\sqrt{6}}\log \left(\frac{H(r)}{K(r)} \right),~
B=B_t(r) dt,~~A^{(I)}=A_t(r) \delta^{I3} dt,
\\
\nonumber
&&{~~~~~~~~~~~~~~~~~~~}{\rm with}~~~
B_t(r)=\Phi^a-\frac{M q}{4 q^2+Mr^2},~A_t(r)=\Phi^A-\frac{MQ}{2Q^2+Mr^2 }.~ 
\end{eqnarray} 
This solution is written in terms of three parameters $(M,Q,q)$, corresponding (up to some factors) to
the global mass and two electric charges. 

In what follows, to avoid cluttering our expressions with complicated factors of $g_{YM}$, 
we use the observation that the above solution is left invariant by the transformation
$r\to \lambda r$, $g_{YM}\to g_{YM}/\lambda$, $(q,Q)\to \lambda (q,Q)$ and $(x,y,z)\to \lambda (x,y,z)$,
and we set $g_{YM}=1$ without any loss of generality. 

The horizon is located $r=r_H$, with $r_H$ the largest positive root of the equation $f(r)=0$, which reduces to
%\begin{eqnarray}
%\label{metricU1xU1-41}
 $
r_H^6+4(\frac{q^2+Q^2}{M})r_H^4+\frac{4}{M^2}(Q^4+4q^2Q^2-2M^3)r_H^2+\frac{16q^2Q^4}{M^3}=0.
$
%\end{eqnarray}
Although one can write an expression for $r_H(M,q,Q)$, it turns out to be more convenient to express $q$ in terms of $r_H,Q,M$:
\begin{eqnarray}
\label{metricU1xU1-9} 
q=\frac{r_H}{2\sqrt{M}}\sqrt{\frac{8M^3}{(2Q^2+Mr_H^2)^2}-1}~.
\end{eqnarray}  

As usual, the constants $\Phi^A,~\Phi^a$ in the expressions of $A_t(r),B_t(r)$ are found by imposing the regularity of the 
one-forms $A,B$ on the horizon, which implies
\begin{eqnarray}
\label{metricU1xU1-7} 
 \Phi^A=\frac{MQ}{2Q^2+Mr_H^2 },~~
  \Phi^a=\frac{M q}{4 q^2+Mr_H^2}.
\end{eqnarray}
Thus, physically they correspond to the two chemical potentials associated with the system.

A straightforward computation leads to the following expressions for the 
mass ${\cal M}$,  electric charges ${\cal Q}_A$ and ${\cal Q}_a$ for the $SU(2)$ and $U(1)$ fields,  
entropy $S$ and Hawking temperature $T_H$:
\begin{eqnarray}
\label{metricU1xU1-10} 
\nonumber
&&
{\cal M}=\frac{3M}{16 \pi} {\cal V},~~
{\cal Q}_A=\frac{1}{2\pi}Q {\cal V},~~
{\cal Q}_a=\frac{1}{2\pi}q {\cal V},~~ 
 S=\frac{1}{2}\sqrt{2M}r_H {\cal V},~~
\\
\nonumber
&&
T_H=\frac{1}{32\sqrt{2}\pi}\frac{ 
8 Q^6+8M^4r_H^2+12 MQ^4r_H^2+6 M^2Q^2r_H^4
+M^3(-16Q^2+ r_H^6)
 }{M^{5/2}r_H(2Q^2+Mr_H^2)},
\end{eqnarray} 
with ${\cal V}=\int d^3x$; however, for the rest of this work, to simplify the expressions,
we set ${\cal V}=1$, $i.e.$ we shall work with mass, entropy and electric charge densities. 

A straightforward computation shows that the solutions satisfy the first law of thermodynamics,
%\begin{eqnarray}
%\label{metricU1xU1-11} 
$
d{\cal M}=T_H dS +\Phi^A d{\cal Q}_A+\Phi^a d{\cal Q}_a,
$
%\end{eqnarray}
and the Smarr law,
%\begin{eqnarray}
%\label{metricU1xU1-12} 
$
 {\cal M}=\frac{3}{4}(T_H  S +\Phi^A  {\cal Q}_A+\Phi^a  {\cal Q}_a).
 $
%\end{eqnarray}

%%%%%%%%%%%%%%%%%%%%%%%%%%%%%%%%%%%%%%%%%%%%%%%%%%%%%%%%%%%%%%%%%%%%%%%%%%%%%%%%%%%%%%%%%
 \subsection{Thermodynamic properties}
%%%%%%%%%%%%%%%%%%%%%%%%%%%%%%%%%%%%%%%%%%%%%%%%%%%%%%%%%%%%%%%%%%%%%%%%%%%%%%%%%%%%%%%%% 
These $U(1)\times U(1)$ solutions possess a relatively complicated thermodynamics.
Restricting for simplicity to a canonical ensemble,   we study black branes holding the temperature
$T_H$, 
and the charges  ${\cal Q}_A,~{\cal Q}_a$  fixed. The associated thermodynamic potential is the Helmholz free energy 
%\begin{eqnarray}
%\label{F}
$F[T_H;{\cal Q}_A,{\cal Q}_a]={\cal M}-T_HS ~.$
%\end{eqnarray}
Thermodynamic stability requires the  positivity
of the specific heat at constant electric charges, $C =T_H(\partial S/\partial T_H)$. A useful relation here is
% \begin{eqnarray}
% \label{metricU1xU1-20}
$
T_H=\frac{1}{\pi 2^{17/3} }
\frac{S^6+(Q^2+q^2)S^4-Q^4 q^2}{S^{5/3}\left((Q^2+S^2)^2(2q^2+S^2)\right)^{2/3}} ,
$
%\end{eqnarray}
which defines $S(T_H;{\cal Q}_A,{\cal Q}_a)$, although an explicit formula cannot be written in the general case.

Analytic results are found 
only when discussing the limiting cases with a vanishing $Q$ or $q$.
The properties of the solutions with $Q=0$ ($i.e.$ a consistent truncation of the model with a 
$U(1)$ field only, $F_{\mu\nu}^{(I)}=0$)
are discussed at length in \cite{Aprile:2010ge}. 
No extremal configurations are found in this case, since the temperature is bounded from below, $T_H^{(min)}>0$.
For any given $T_H>T_H^{(min)}$, there are two branches of solutions, one of them being thermally stable.

By constrast, the solutions with $q=0$ ($SU(2)$ gauge fields only, $f_{\mu \nu}=0$),
admit an extremal limit which is approached for $M=Q^{4/3}/2^{1/3}$.
The entropy of the extremal solutions vanishes, a number of invariant quantities diverging in that limit.
For nonextremal configurations one finds a single branch of solutions, with
 \begin{eqnarray}
 \label{metricU1xU1-23}
\nonumber
&
S=4\pi T_H Q^{2/3} 
\bigg [
\frac{32\pi^2 T_H^2}{3Q^{2/3}}+
\bigg(1+\frac{32768\pi^6T_H^6}{27Q^2}-\sqrt{1+\frac{65536 \pi^6T_H^6}{27Q^2}} \bigg)^{1/3}
%\\
%\nonumber
+
\bigg(1+\frac{32768\pi^6T_H^6}{27Q^2}+\sqrt{1+\frac{65536 \pi^6T_H^6}{27Q^2}}\bigg)^{1/3}
\bigg ],
\end{eqnarray}
which possesses a positive specific heat. 

The  solutions with two $U(1)$ charges exhibit a complicated picture, which is governed by the value of the relative ratio  $q/Q$.
The picture in Figure 1 appears to be generic: for any $Q\neq 0$,
the solutions with small 

%%%%%%%%%%%%%%%%%%%%%%%%%%%%%%%%%%%%%%%%%%%%%%%%%%%%%%%%%%%%%%%%
\vspace*{-0.1cm}
 {\small \hspace*{3.cm}{\it  } }
\begin{figure}[ht]
\hbox to\linewidth{\hss%
	\resizebox{8cm}{6cm}{\includegraphics{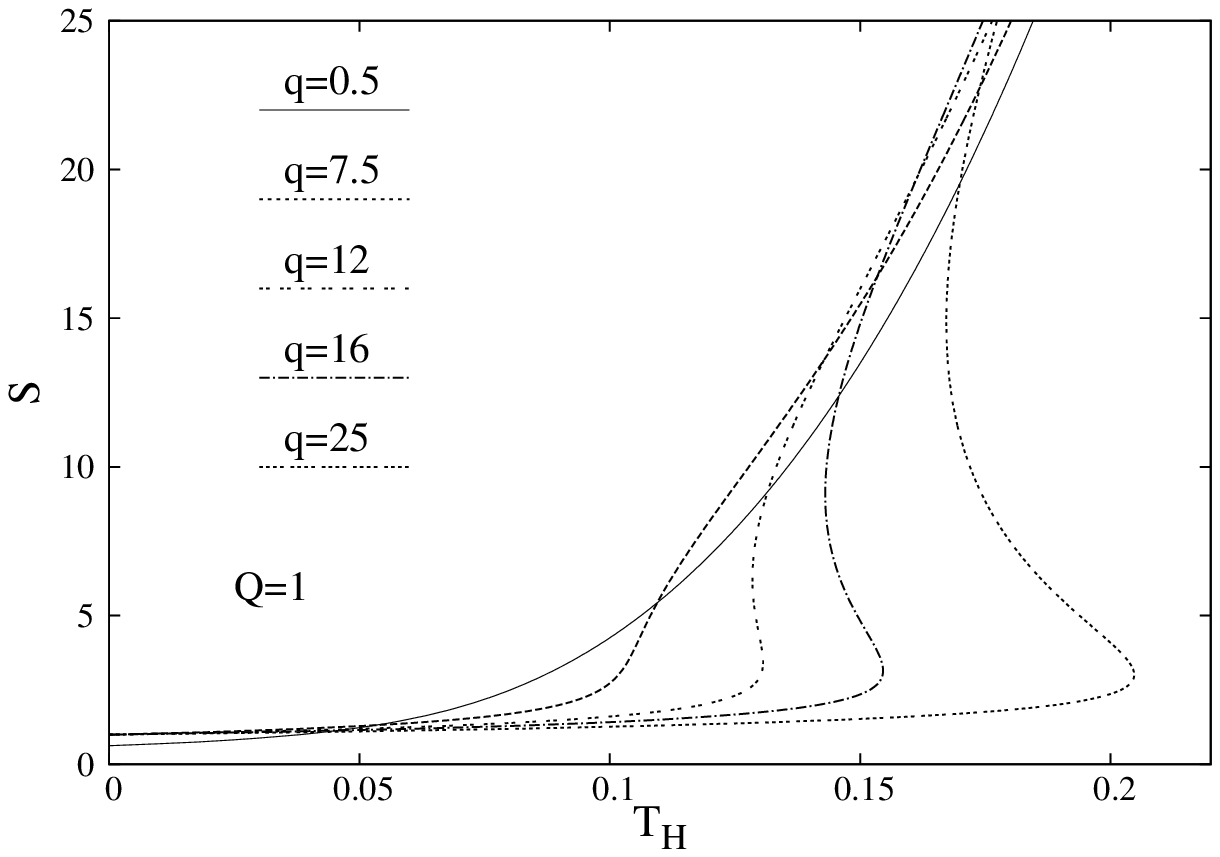}}
\hspace{10mm}%
        \resizebox{8cm}{6cm}{\includegraphics{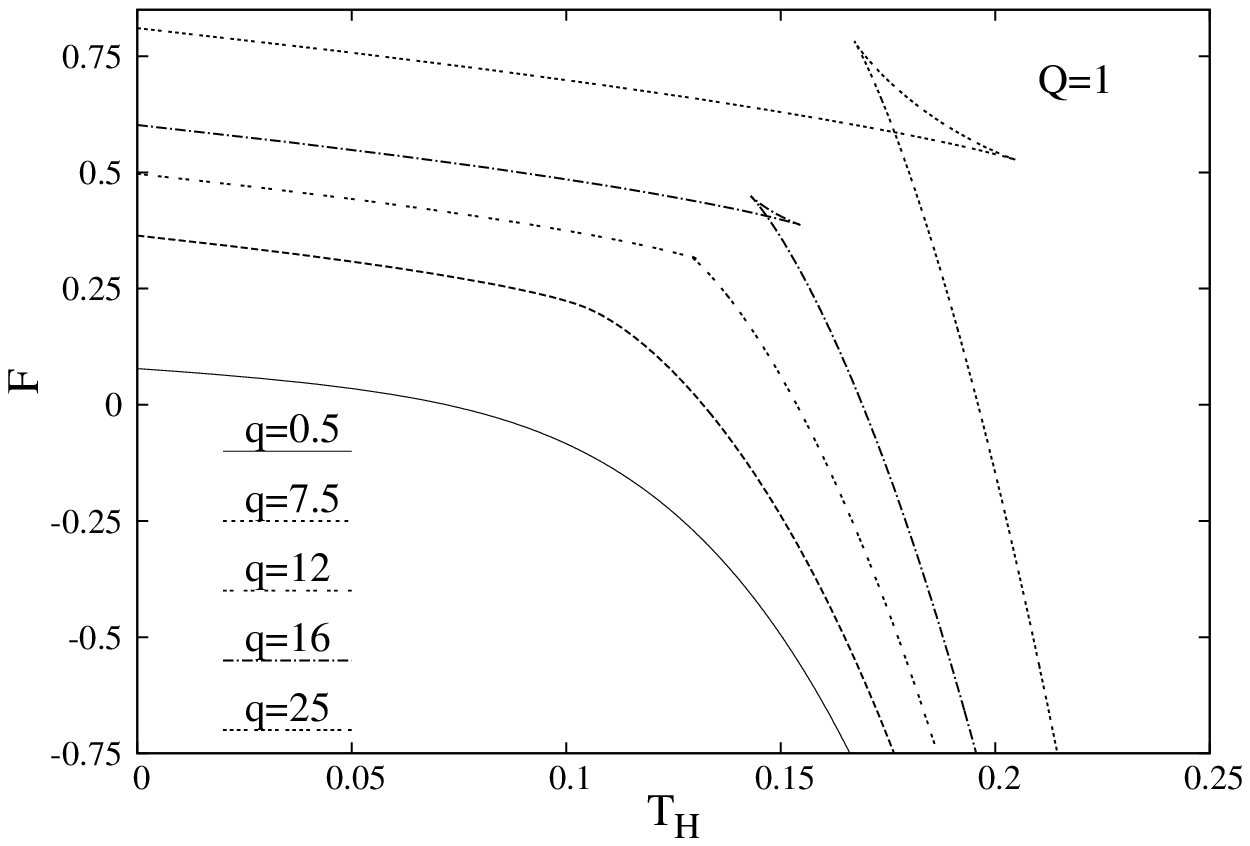}}	
\hss}
 \caption{\small 
The entropy and free energy of the Abelian black brane solutions is shown
for several values of the electric charges. 
  } 
\end{figure}
%%%%%%%%%%%%%%%%%%%%%%%%%%%%%%%%%%%%%%%%%%%%%%%%%%%%%%%%%%%%%%%%%%%%%%%
\\
enough $q$ are thermally stable, the entropy increasing with the temperature (note that $S(T_H=0)\neq 0$, 
the geometry remaining regular in this limit).

However, when increasing $q$ we notice the occurance of three branches of solutions for some intermediate
 range of $T_H$. 
The physically relevant branch (which has less free energy) is the third one, which continues to  $T_H\to \infty$
(the large temperature behaviour is $S=128 \pi^3 T_H^3 +O(T_H)$).
Also, the second branch is unstable since it possess a negative specific heat.

For a more systematic discussion of the properties of the generic Abelian solutions, it turns out convenient to
work with the following scaled quantities\footnote{The $U(1)\times U(1)$ exact solution 
has an extra scaling symmetry 
${\cal M}\to \lambda^4 {\cal M}$, 
${\cal Q}_k\to \lambda^3 {\cal Q}_k$, 
$T_H\to \lambda  T_H$,
$S\to \lambda^3  S$ and
$\Phi^k\to \lambda  \Phi^k$,
with $k=(a,A)$ and $\lambda>0$
an arbitrary constant.
The quantities in (\ref{metricU1xU1-13})
are left invariant by this transformation.
 } 
\begin{eqnarray}
\label{metricU1xU1-13}  
q_A=c_1 \frac{{\cal Q}_A}{{\cal M}^{3/4}},
~~
q_a=c_4  \frac{{\cal Q}_a}{{\cal M}^{3/4}},
~~
s=c_2  \frac{{\cal S}}{{\cal M}^{3/4}},
~~
t_H=c_3 \frac{T_H}{{\cal M}^{1/4}},
~~
\varphi^A=c_5 \frac{\Phi^A}{{\cal M}^{1/4}},
~~
\varphi^a=c_6 \frac{\Phi^a}{{\cal M}^{1/4}},
\end{eqnarray}
with 
$c_1=\frac{3^{3/4}}{4}(\frac{\pi}{2})^{1/4}$,
$c_2=\frac{1}{2^{13/4}}(\frac{3}{\pi})^{3/4}$,
$c_3=4\times 6^{1/4}\pi^{3/4}$,
$c_4=\frac{3^{3/4}}{2\sqrt{2}}(\frac{\pi}{2})^{1/4}$,
$c_5= (\frac{6}{\pi})^{1/4}$, and
$c_6= 2^{3/4}(\frac{3}{\pi})^{1/4}$ 
being constant factors which have been chosen such that the expressions below take a simpler form.

All the relevant quantities can then  be expressed in terms of $q_A$ and $s$ only: 
\begin{eqnarray}
\label{metricU1xU1-14} 
t_H=\frac{(q_A^2+s^2)^3+s^2-q_A^2}{s(q_A^2+s^2)},~
 q_a=s\sqrt{\frac{1}{(q_A^2+s^2)^2}-1}, ~~
\varphi^A=\frac{q_A}{q_A^2+s^2},~~
\varphi^a=\frac{(q_A^2+s^2)^2}{s}\sqrt{\frac{1}{(q_A^2+s^2)^2}-1}.~~{~}
\end{eqnarray} 
It is clear that all solutions satisfy the condition 
\begin{eqnarray}
\label{c1}
\nonumber 
s^2 \leq 1-q_A^2,
\end{eqnarray}
the upper bound being approached for solutions with $SU(2)$ fields only.
Moreover, the condition $t_H\geq 0$ imposses a lower bound for the reduced entropy:
\begin{eqnarray} 
\label{metricU1xU1-17}
s^2\geq U(q_A)-q_A^2,
\end{eqnarray}
 where
\begin{eqnarray} 
\label{metricU1xU1-191}
 U(q_A)=\bigg(\sqrt{\frac{1}{27}+q_A^4} +q_A^2\bigg)^{1/3}-\bigg(\sqrt{\frac{1}{27}+q_A^4} -q_A^2\bigg)^{1/3}.
\end{eqnarray}
One can also show that the scaled $U(1)$ charge $q_a$ has a finite range, with 
\begin{eqnarray} 
\label{metricU1xU1-18}
0\leq q_a^2\leq \frac{\left(1-U^2(q_A)\right)^2}{2U(q_A)}.
\end{eqnarray}
Solutions with a maximal value of $q_a$ correspond to extremal black holes,
with $t_H=0$ and $s^2= U(q_A)-q_A^2$. 
From (\ref{metricU1xU1-14}), the reduced entropy of the extremal solutions can also be written as
\begin{eqnarray} 
\label{metricU1xU1-19}
 s^2=\frac{1}{4}\left(-q_a^2+q_a\sqrt{q_a^2+8q_A^2} \right),
\end{eqnarray}
which is a nonvanishing quantity for $q_A\neq 0$.

%%%%%%%%%%%%%%%%%%%%%%%%%%%%%%%%%%%%%%%%%%%%%%%%%%%%%%%%%%%%%%%%%%%%%%%%%%%%%%%%%%%%%%%%%
\section{The superfluid phase}
%%%%%%%%%%%%%%%%%%%%%%%%%%%%%%%%%%%%%%%%%%%%%%%%%%%%%%%%%%%%%%%%%%%%%%%%%%%%%%%%%%%%%%%%% 
It is clear that the $U(1)\times U(1)$ solutions should possess non-Abelian generalizations.
These configurations are found when enlarging the $SU(2)$ ansatz to include a nonzero magnetic potential such that
the gauge potential  $A^{(I)}=A_t(r) \delta^{I3} dt$ is approached only asymptotically.

Following previous work  \cite{Manvelyan:2008sv}, \cite{Ammon:2009xh}
on pure Einstein-Yang-Mills (EYM) solutions with vector hair, we choose 
an $SU(2)$ gauge fields ansatz  possessing both electric and magnetic potentials,
while the $U(1)$ ansatz is still purely electric:
 \begin{eqnarray}
\label{SU2-an}
A^{(I)}=w(r)\delta^{I1}dx+A_t(r) \delta^{I3} dt ,~~B=B_t(r) dt.
 \end{eqnarray} 
Also, as before, the dilaton field will depend on the $r$-coordinate only. 
 This leads to a diagonal energy-momentum tensor  and thus it is
 consistent to choose a diagonal metric ansatz.  
 
%%%%%%%%%%%%%%%%%%%%%%%%%%%%%%%%%%%%%%%%%%%%%%%%%%%%%%%%%%%%%%%%%%%%%%%%%%%%%%%%%%%%%%%%%
\subsection{Zero modes for the $U(1)\times U(1)$ black brane}
%%%%%%%%%%%%%%%%%%%%%%%%%%%%%%%%%%%%%%%%%%%%%%%%%%%%%%%%%%%%%%%%%%%%%%%%%%%%%%%%%%%%%%%%%  

Before discussing the general solutions, it is instructive to consider the perturbative limit of
the problem.
Then $w(r)$ is treated as a small perturbation around the  $U(1)\times U(1)$ solutions, $w(r)=\epsilon W(r)$.
After substituting into the linearized YM equations, one finds that $W(r)$ solves
 \begin{eqnarray}
\label{pert2}
W''+ (\frac{1}{r}-\frac{K'}{K}+\frac{f'}{f})W'+\frac{A_t^2 H^2K}{f}W=0.
\end{eqnarray}
For   $Q=0$ one finds the following exact solution of the above equation
\begin{eqnarray}
\label{pert1}
 W(r)=c_0+c_1
 \left (
 \log (1-(\frac{r_H}{r})^2)-\frac{r_H^8}{64M^2}\log (1+\frac{8M}{r_H^2r^2}) 
 \right),
\end{eqnarray}
(where $c_0,c_1$ are arbitrary constants). As one can see, this solution posseses an essential
logarithmic singularity at the horizon and thus cannot be treated as a perturbation.
Thus we conclude that only solutions which are charged with respect the $SU(2)$ fields may posses an instability. 
 
 Although for $Q\neq 0$ the equation (\ref{pert2}) does not appear to be solvable in terms
of known functions, one can construct approximate solutions near the horizon and at infinity. 
As $r\to r_H$, the function $W(r)$ behaves as 
$
W(r)=b+ O(r-r_H)^2,
$
 while, for large $r$,   the approximate form of $W(r)$ is
$
W(r)= {J}/{r^2}+O(1/r^4),
$
with $b$ and $J$ free parameters.  
 Solutions interpolating between these asymptotics  
are constructed numerically\footnote{In this work we restrict our 
study to solutions with a monotonic behaviour of $W(r)$.}.

The mechanism triggering the instability is similar to the pure EYM-$\Lambda$ case \cite{Gubser:2008zu},
the magnetic gauge potential acquiring a tachyonic mass term  near the horizon.
Interestingly, the picture found for $q=0$ is rather similar to that valid for $Q=0$
since in this case too no solutions of (\ref{pert2}) with correct asymptotics are found.
We conclude that, somehow unexpectedly, both electric charges (associated with the 
$SU(2)$ and $U(1)$ fields) should be
nonvanishing for the existence of a normalizable zero mode.

%%%%%%%%%%%%%%%%%%%%%%%%%%%%%%%%%%%%%%%%%%%%%%%%%%%%%%%%%%%%%%%%
\vspace*{-0.1cm}
 {\small \hspace*{3.cm}{\it  } }
\begin{figure}[ht]
\hbox to\linewidth{\hss%
	\resizebox{8cm}{6cm}{\includegraphics{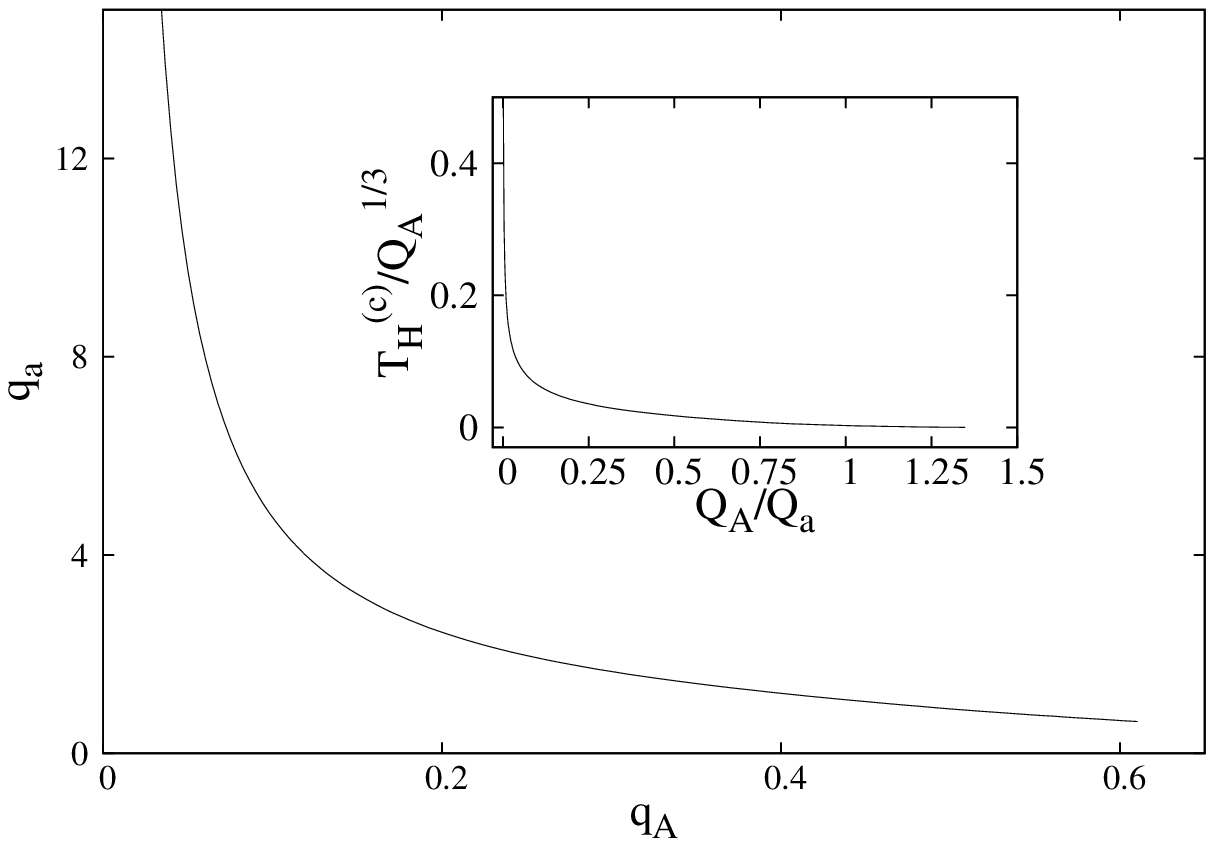}}
\hspace{10mm}%
        \resizebox{8cm}{6cm}{\includegraphics{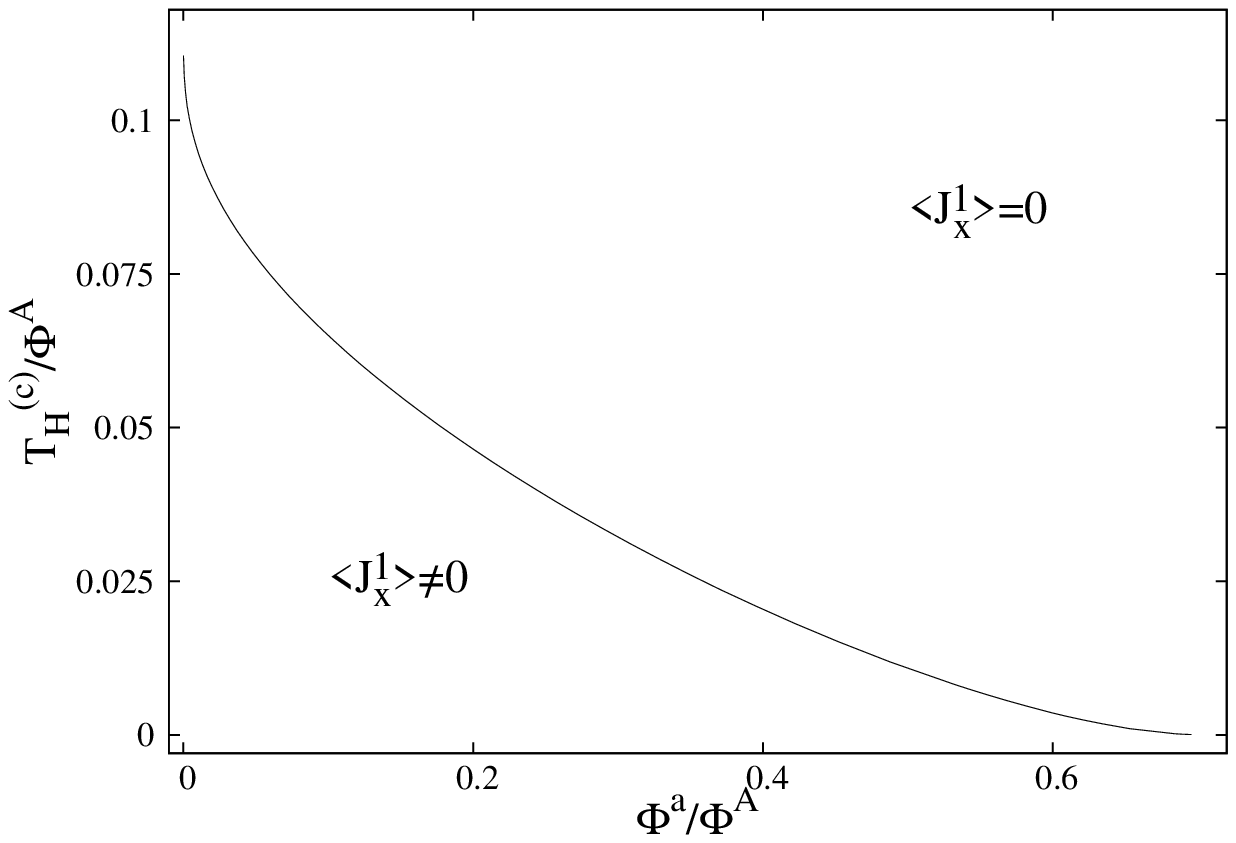}}	
\hss}
 \caption{\small 
Critical curves in the parameter space  where static linear normalizable non-Abelian
perturbations  arise.  } 
\end{figure}
%%%%%%%%%%%%%%%%%%%%%%%%%%%%%%%%%%%%%%%%%%%%%%%%%%%%%%%%%%%%%%%%%%%%%%%

 Some results of the numerical integration are shown in Figure 2.
There, the part of the parameter space above the curves corresponds to
the unbroken phase, where only Abelian solutions exist.
In Figure 2 (left) we show the critical curve in the $(q_A,q_a)$
plane corresponding to configurations unstable with respect to non-Abelian perturbations.
One can see that the reduced $SU(2)$ charge $q_A=c_1 {{\cal Q}_A}/{{\cal M}^{3/4}} $ has a finite range, $0<q_A<0.618$ ,
an extremal configuration (with $T_H^{(c)}\to 0$) being approached for the maximal value of $q_A$ and $q_a\to 0.629$ (corresponding to $\Phi^a/\Phi^A \simeq  0.704$).
In Figure 2 (right) we show the critical temperature  $T_H^{(c)}$ as a function of the $U(1)$ chemical
potential $\Phi^a$ (both quantities are normalized $w.r.t.$ the  $SU(2)$ chemical
potential $\Phi^A$);
note that the critical temperature is monotonically decreasing
 as we increase the ratio $\Phi^a/\Phi^A$.
  
%%%%%%%%%%%%%%%%%%%%%%%%%%%%%%%%%%%%%%%%%%%%%%%%%%%%%%%%%%%%%%%%%%
\subsection{Black holes with non-Abelian hair}
%%%%%%%%%%%%%%%%%%%%%%%%%%%%%%%%%%%%%%%%%%%%%%%%%%%%%%%%%%%%%%%%% 
 
%%%%%%%%%%%%%%%%%%%%%%%%%%%%%%%%%%%%%%%%%%%%%%%%%%%%%%%%%%%%%%%%%%
\subsubsection{The equations and global charges}
%%%%%%%%%%%%%%%%%%%%%%%%%%%%%%%%%%%%%%%%%%%%%%%%%%%%%%%%%%%%%%%%% 
The instability of the  $U(1)\times U(1)$  
solution pointed out in the previous section can be viewed as an indication 
of the existence of a branch of non-Abelian solutions with nontrivial
magnetic non-Abelian fields outside the horizon.

In the numerical construction of these solutions, we adopt the following metric
ansatz\footnote{The line element  (\ref{metricU1xU1-1}) of the 
$U(1)\times U(1)$ solution  can also be written in the form (\ref{metric-an}) (with $f(r)=1$)
 by defining a new radial coordinate.
 However, this results in rather complicated expressions.}, 
which was first proposed in \cite{Manvelyan:2008sv} for the case of the pure EYM-$\Lambda$ system:
\begin{eqnarray}
\label{metric-an}
ds^2=\frac{dr^2}{N(r)}+r^2 \left ( \frac{dx^2}{f^4(r)}+f^2(r)(dy^2+dz^2)\right)
-N(r)\sigma^2(r)dt^2,~~{\rm with}~~N(r)=-\frac{4m(r)}{3r^2}+\frac{r^2}{\ell^2}.
 \end{eqnarray}
Plugging (\ref{metric-an}) and (\ref{SU2-an})   into (\ref{Einstein-eqs})-(\ref{Tij})  results in the equations of
motion\footnote{One can see that  the Chern-Simons  term in (\ref{action5}) does not contribute to
 the equations of motion so that the gauge fields do not interact directly.}:
 \begin{eqnarray} 
\nonumber
&&m'=\frac{3r^3Nf'^2}{2f^2}
+\frac{r e^{2 a \phi}}{2g_{YM}^2}
\bigg(
f^4 Nw'^2+\frac{r^2A_t'^2}{\sigma^2}+\frac{f^4A_t^2w^2}{N\sigma^2}
\bigg)
\\
\label{eq-num}
&&{~~~~~~}+\frac{e^{-4a\phi}r^3B_t'^2}{2\sigma^2}
+\frac{1}{2}r^3N\phi'^2
+\frac{g_{YM}^2}{8}r^3
(3g_{YM}^2-e^{-2a \phi}-2 e^{a\phi}),
\\
\nonumber
&&\frac{\sigma'}{\sigma}=\frac{2r}{3}
\left(
\frac{3f'^2}{f^2}
+\phi'^2
+\frac{1}{g_{YM}^2}\frac{e^{2a\phi}}{r^2}
(f^4w'^2
+\frac{A_t^2w^2f^4}{N^2\sigma^2}
\right),
 \end{eqnarray}
 \begin{eqnarray}
\label{eqs2}
\nonumber
&&(r^3N\sigma\frac{f'}{f^2})'=\frac{2 e^{2a\phi}}{3g_{YM}^2}r\sigma f^3N
(w'^2-\frac{A_t^2w^2}{N^2\sigma^2})
-r^3N\sigma\frac{f'^2}{f^3},
\\
\nonumber
&&(r^3N\sigma\phi')'=ar\sigma
\bigg[
\frac{ e^{2a\phi}}{g_{YM}^2}\left(f^4Nw'^2-\frac{r^2A_t'^2}{\sigma^2}-\frac{f^4A_t^2w^2}{N\sigma^2} \right)
%\\
%\nonumber
 +\frac{2 e^{-4 a \phi}}{\sigma^2}r^2B_t'^2+\frac{g_{YM}^2}{8}(e^{-2 a\phi}-e^{a\phi})r^2
\bigg],
\\
\nonumber
&&(e^{2a\phi}r^3\frac{A_t'}{\sigma})'=\frac{e^{2a\phi}rf^4}{N\sigma}w^2A_t,
~~ (e^{-4a\phi}r^3\frac{B_t'}{\sigma})'=0,~~
%\\
%\nonumber
%&&
 (e^{2 a \phi} N f^4 r \sigma w' )' = - \frac{e^{2 a \phi} r f^4}{N \sigma} A_t^2 w \ .
\end{eqnarray}
These equations possess the following scaling symmetries
(invariant functions are not shown)
\begin{eqnarray}
\label{scal1}
&&(i)~ \sigma \to \lambda \sigma,~~A_t\to \lambda A_t,~~B_t\to \lambda B_t,~~~~~~~~~~~
%\\
%\label{scal1} 
 (ii)~f \to \lambda f,~~w\to \frac{w}{\lambda^2},
\\
\nonumber
&&(iii)~r\to \lambda r,~~g_{YM}\to \frac{g_{YM}}{\lambda},~~A_t\to \frac{A_t}{\lambda},~~~~~~
%\\
%\nonumber
 (iv)~r\to \lambda r,~m\to \lambda^4 m,~A_t\to \lambda A_t,~B_t\to \lambda B_t,~w\to \lambda w,
\end{eqnarray}
with $\lambda>0$ an  arbitrary number.
The symmetries $(i)$ and $(ii)$ are used to set $\sigma(\infty)=1$, $f(\infty)=1$, 
while $(iii)$ is used to set $g_{YM}=1$ without any loss of generality, which
 fixes the AdS length scale, $\ell=2\sqrt{2}$.
 Note also that the last equation in (\ref{eq-num}) implies the existence of the first integral
\begin{eqnarray}
\label{fi}
B_t'=\frac{2e^{4a\phi}\sigma q}{r^3}, 
\end{eqnarray}
with $q$ a constant fixing the $U(1)$ electric charge.

We consider again black branes with an horizon at $r=r_h$, where $N(r_h)=0$.
The non-extremal solutions 
have the following expansion as $r\to r_H$:
\begin{eqnarray}
\nonumber
&&
m(r)=\frac{3}{4}\frac{r_H^4}{\ell^2}+ O(r-r_H) ,~~
\sigma(r)=\sigma_h+ O(r-r_H) ,~~f(r)=f_h+ O(r-r_H)^2,~~\phi(r)=\phi_0+ O(r-r_H) ,
\\
\label{eh-an0}
&&
w(r)=w_h+ O(r-r_H)^2,~  
A_t(r)=V_1(r-r_H)+O(r-r_H)^2,~~
B_t(r)=v_1(r-r_H)+O(r-r_H)^2,~~{~~~~~~}
\end{eqnarray}
with the independent parameters $\{\sigma_h,~f_h,~\phi_0,~w_h,~v_1,~V_1 \}$
which fix  the  coefficients of all higher order terms.

We are interested in solutions approaching the $U(1)\times U(1)$ configurations asymptotically.
We assume\footnote{The generic solutions
have a more complicated asymptotic behaviour, with 
$\omega(r)=\omega_0-\ell^2 \omega_0^3  \frac{\log r}{r^2}+\dots$,
$\phi(r)=\frac{\alpha}{r^2}+\beta\frac{ \log r}{r^2}+\dots,$
which implies the existence of $\log$ terms also in the expression of the metric functions,
$e.g.$ $m(r)=M+ \frac{1}{16}\beta(\beta-4\alpha) \log r-\frac{\beta^2}{8}\log^2 r
+\frac{3}{2}\omega_0^4 \log r+\dots$~.} that, as $r\to \infty$, $w(r)$ vanishes and $\phi(r)$ decays as $1/r^2$.
A systematic analysis then reveals the following expansion of the solutions
at large $r$:
\begin{eqnarray}
\nonumber
&&
m(r)=M+O(1/r^2),~~
\sigma(r)=1-\frac{2}{3}\frac{\alpha^2}{r^2}+ O(1/r^4),~~
\phi(r)=\frac{D}{r^2}+ O(1/r^4),~~
f(r)=1+\frac{f_4}{r^4}+ O(1/r^6),~~
\\
\label{bc23} 
&&
w(r)=\frac{J}{r^2}+ O(1/r^4),~~A_t(r)=\Phi^A-\frac{Q}{r^2},~~B_t(r)=\Phi^a-\frac{q}{r^2},
\end{eqnarray}
with  $\{M,J,Q,q,D,f_4,\Phi^A,\Phi^a\}$ arbitrary coefficients. 

All physical quantities are fixed by the data at the horizon and at infinity.
As in the Abelian case, the global charges are the mass and the $SU(2)$ and $U(1)$
electric charges, with\footnote{Note that, different from the pure EYM$-\Lambda$ case, the total
mass is not given by the asymptotic value of $m(r)$, aquiring a contribution from the scalar field.} 
\begin{eqnarray}
{\cal M}=\frac{1}{4 \pi}(M+\frac{D^2}{\ell^2}),~~
{\cal Q}_A=\frac{1}{2\pi}Q ,~~
{\cal Q}_a=\frac{1}{2\pi}q ,~~
\end{eqnarray} 
while $\Phi^A$, $\Phi^a$ are chemical potentials associated with the two gauge fields.
The entropy and Hawking temperature of the solutions are given by
\begin{eqnarray}
S=\frac{1}{4}r_H^3,~~T_H=\frac{\sigma_h}{2\pi}
\left [
\frac{r_H}{3}(\frac{1}{2}(2e^{a\phi_0}+e^{-2a\phi_0})
-\frac{2e^{2a\phi_0}v_1^2}{\sigma_h^2})-\frac{8e^{4a\phi_0}q^2}{3r_h^5}
\right ].
\end{eqnarray}
%%%%%%%%%%%%%%%%%%%%%%%%%%%%%%%%%%%%%%%%%%%%%%%%%%%%%%%%%%%%%%%%
\vspace*{-0.1cm}
 {\small \hspace*{3.cm}{\it  } }
\begin{figure}[ht]
\hbox to\linewidth{\hss%
	\resizebox{8cm}{6cm}{\includegraphics{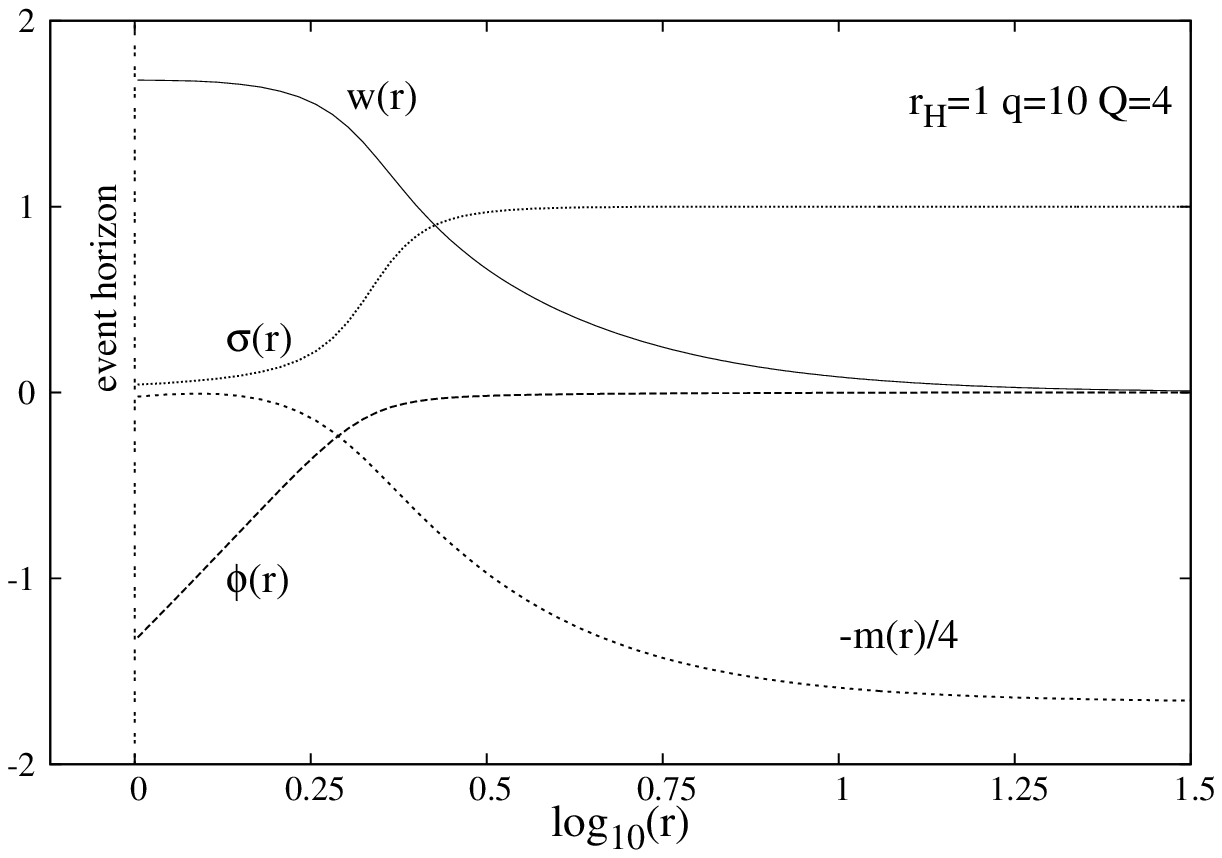}}
\hspace{10mm}%
        \resizebox{8cm}{6cm}{\includegraphics{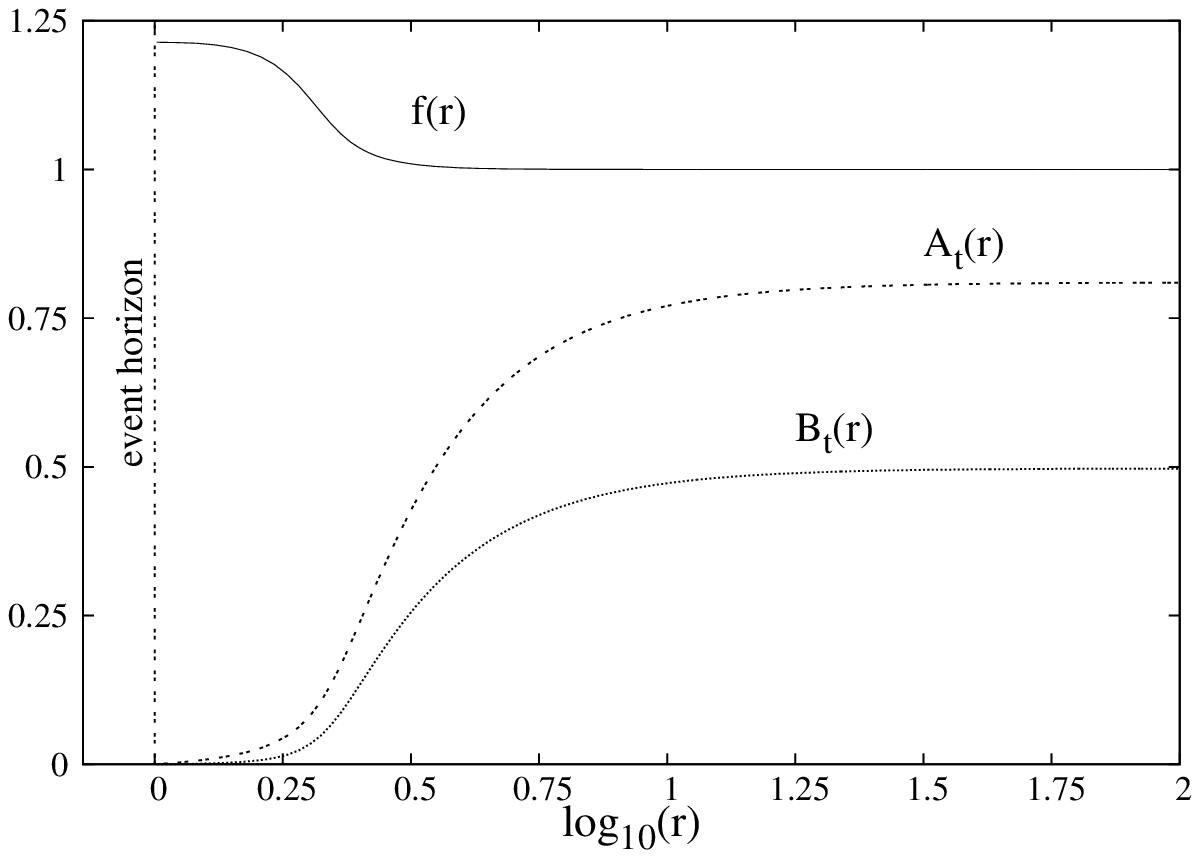}}	
\hss}
 \caption{\small 
The profile  of a typical non-Abelian solution is shown as a function of the radial coordinate.  } 
\end{figure}
%%%%%%%%%%%%%%%%%%%%%%%%%%%%%%%%%%%%%%%%%%% 

For completness, we mention that the boundary stress-tensor $\mathsf{T}_i^j$
as defined by (\ref{s1})
is diagonal, with the nonzero components:
 \begin{eqnarray}
\mathsf{T}_x^x=\frac{1}{8\pi}\frac{2}{3\ell}(D^2-12 f_4+M\ell^2),~~
\mathsf{T}_y^y=\mathsf{T}_z^z=\frac{1}{8\pi}\frac{2}{3\ell}(D^2+6 f_4+M\ell^2),~~
\mathsf{T}_t^t=-\frac{1}{8\pi }\frac{2}{ \ell}(D^2+M\ell^2),~~
\end{eqnarray}
such that $\mathsf{T}_i^i=0$.

%%%%%%%%%%%%%%%%%%%%%%%%%%%%%%%%%%%%%%%%%%%%%%%%%%%%%%%%%%%%%%%%%%
\subsubsection{Numerical solutions}
%%%%%%%%%%%%%%%%%%%%%%%%%%%%%%%%%%%%%%%%%%%%%%%%%%%%%%%%%%%%%%%%%
%
Eqs. (\ref{eq-num}) with boundary conditions (\ref{eh-an0}) and (\ref{bc23}), respectively, have been solved numerically
using a standard shooting method. In addition to using this algorithm, some solutions were also constructed by 
employing a collocation method for boundary-value ordinary
differential equations equipped with an adaptive mesh selection procedure.
% \cite{colsys}.
We have confirmed that there is good agreement between the results obtained with these two different methods. 

As expected, some basic properties of these black branes are rather similar to those found in \cite{Manvelyan:2008sv}, \cite{Ammon:2009xh}
in the case of the purely EYM-$\Lambda$ model.
However, the solutions in the present work feature a second control parameter, which is the $U(1)$ 
electric charge $q_a$ (or equivalently, the chemical potential $\Phi^a$).

For all solutions, the functions $\sigma(r)$ and $A_t(r),B_t(r)$ always increase monotonically
with growing $r$. However, $m(r),f(r),\phi(r)$ and $w(r)$ may feature a more complicated behaviour, with local extrema.
For sufficiently small $\omega_h$, all field variables remain
close to their values for the Abelian configuration with the same $(r_H,Q,q)$.
Significant differences occur for large enough values of $\omega_h$ 
and the effect of the magnetic fields on the geometry becomes increasingly more pronounced.
The profiles of a typical solution illustrating these features are presented in Figure 3.

In the numerical approach, we make use of the existence of the first 
integral (\ref{fi}) to fix the value of the electric charge associated
with the $U(1)$ field, which implies $v_1= {8e^{4a\phi_h}q\sigma_h}/{r_H^3}$ in the   near  horizon expansion (\ref{eh-an0}). 
The scaling symmetry $(iv)$ in (\ref{scal1}) is used to set $r_H=1$,
such that the only remaining control parameters are $w(r_H)$ and $V_1$.

We have studied in a systematic way families of solutions with fixed values of $q$
between 0.5 and 7, the following picture being generic.
First, the behaviour of solutions for arbitrary data on the horizon is such that at large $r$ one finds 
$w\to w_0\neq 0$ and $\phi(r)\to \log r/r^2$
(in which case the total mass as defined according to (\ref{Mcons}) diverges), or else there is a singularity at finite $r$.
Given ($w_h,q;r_H$), solutions with the correct asymptotic behaviour\footnote{These solutions here are also indexed
by the node number of the magnetic potential $w(r)$.  It turns out that
the configurations with nodes represent excited states whose energy is always greater
than the energy of the corresponding nodeless configurations, and are
therefore ignored in what follows.}
 are found only for a discrete set of values of $(V_1$, $\phi_0)$.
 Also, all solutions possess a \\

%%%%%%%%%%%%%%%%%%%%%%%%%%%%%%%%%%%%%%%%%%%%%%%%%%%%%%%%%%%%%%%%%%%%%%%%%%%
\vspace*{-0.2cm}
 {\small \hspace*{3.cm}{\it  } }
\begin{figure}[ht]
\hbox to\linewidth{\hss%
	\resizebox{8cm}{6cm}{\includegraphics{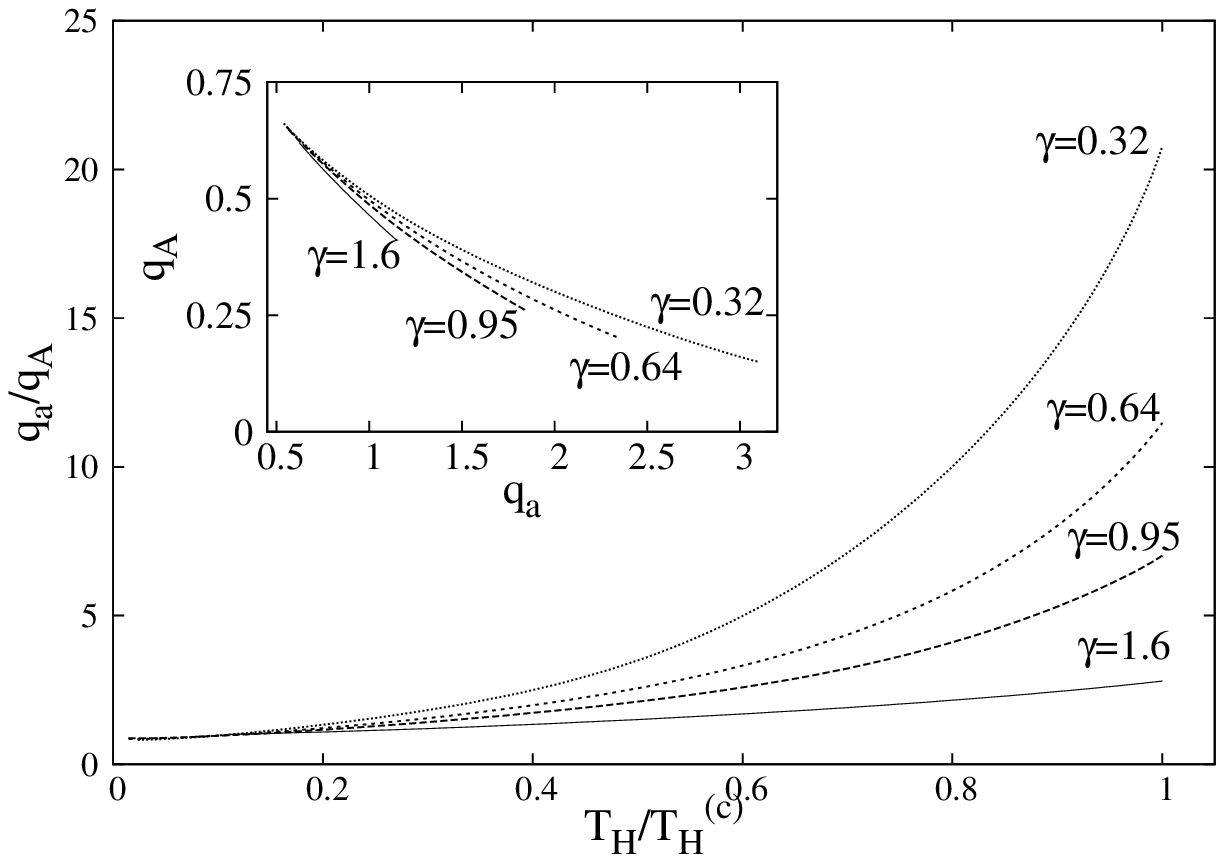}}
\hspace{10mm}%
        \resizebox{8cm}{6cm}{\includegraphics{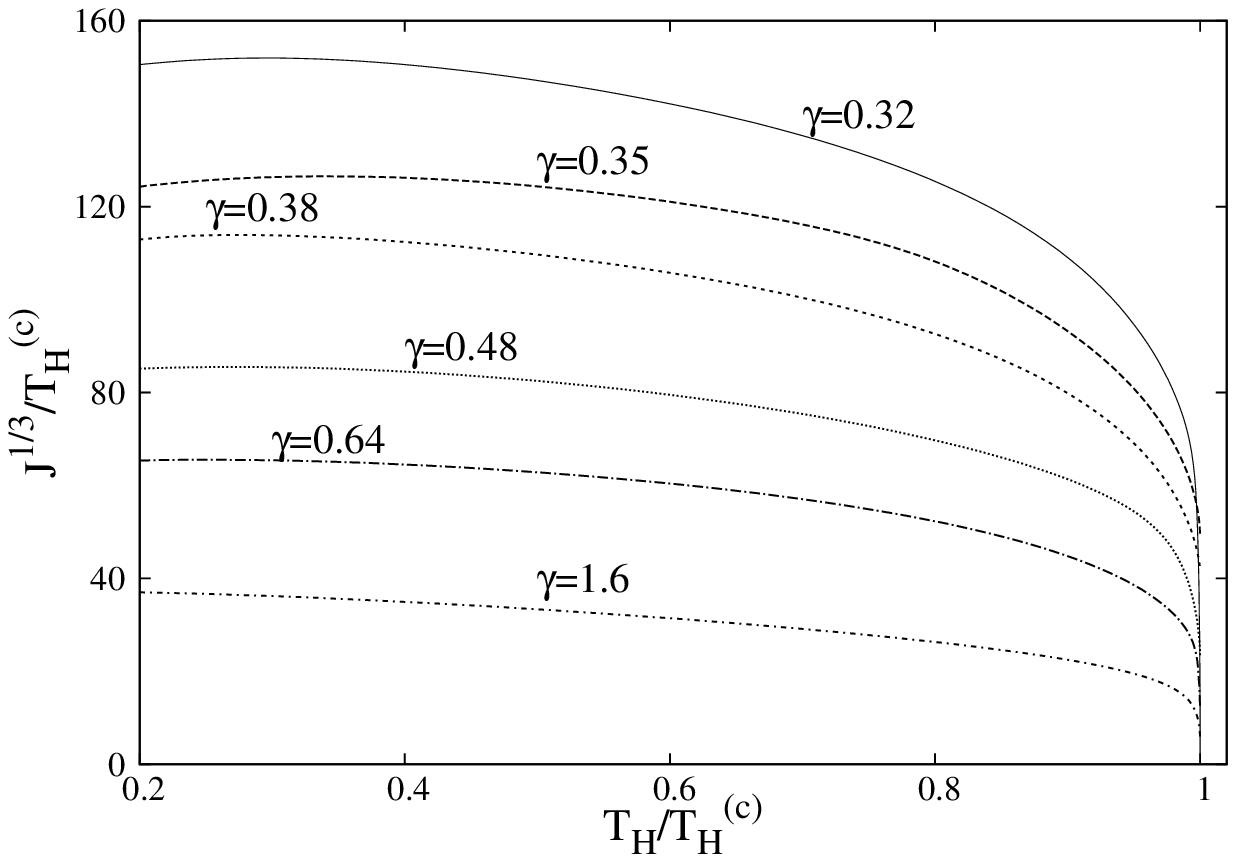}}	
\hss}
 \caption{\small 
Various parameters of the non-Abelian solutions are shown for several
values of of the scale-free ratio $\gamma={\cal Q}_a/S$.  } 
\end{figure}
%%%%%%%%%%%%%%%%%%%%%%%%%%%%%%%%%%%%%%%%%%%%%%%%%%%%%%%%%%%%%%%%%%%%%%%%%%%%%%%%%
\\
non-vanishing electric $Q_A$ associated 
 with the $SU(2)$ field.

Moreover, for fixed $(r_H,q)$, one finds a branch of non-Abelian solutions for $0<w_h<w_h^{max}$.
Along this branch, the Hawking temperature decreases, an extremal configuration being approached\footnote{Our numerical code usually provided good quality solutions
for $T_H\gtrsim T_H^{(c)}/10$.} for the maximal  value of $w_h$.
The numerical construction of the solutions with $T_H=0$ requires a different metric 
ansatz than (\ref{metric-an}) and is beyond the scope of this work.
However, based on the results in  the near-extremal case, we expect the extremal solutions to share the basic properties
of their general Abelian counterparts, possesing a regular horizon with non-vanishing entropy.

%In visualizing the global data, one should work with quantities
%which are invariant under the scaling transformation $(iv)$ in (\ref{scal1}). 

Some results of the numerical integration are shown in Figure 4.
There we employ scale-free quantities defined in (\ref{metricU1xU1-13}),
which are invariant under the scaling transformation $(iv)$ in (\ref{scal1});
also, we have found it convenient to define $\gamma={\cal Q}_a/S$ as a second scale-free control parameter. 
One can see that, for all cases we considered, non-Abelian solutions
exist only for values of the Hawking temperature smaller than a critical temperature
$T_H^{(c)}$.
This is the temperature at which the $U(1)\times U(1)$ solution admits a static
linearised perturbation, with nonvanishing but infinitesimally small $w$.  
Moreover, the dependece of the order parameter $J$ on the Hawking temperature is similar 
to that found in the literature for the $\gamma=0$ case ($i.e.$ a EYM-$\Lambda$ model).
Also, as expected, we have found that the difference in the free energy density, $F$, 
between a non-Abelian solution and the $U(1)\times U(1)$ solution with the same temperature and electric charges
is negative, and thus the non-Abelian solution is thermodynamically favoured.

%%%%%%%%%%%%%%%%%%%%%%%%%%%%%%%%%%%%%%%%%%%%%%%%%%%%%%%%%%%%%%%%%%%%%%%%%%%%%%%%%%%%%%%%%
\section{Further remarks. }
%%%%%%%%%%%%%%%%%%%%%%%%%%%%%%%%%%%%%%%%%%%%%%%%%%%%%%%%%%%%%%%%%%%%%%%%%%%%%%%%%%%%%%%%%
 
In this paper we have studied electrically charged black branes of the 
${\cal{N}}=4^+$ $SU(2)\times U(1)$ gauged supergravity model 
with AdS$_5$ asymptotics. 
Apart from the Abelian $U(1)\times U(1)$ configurations, 
we have given numerical evidence that this model
possesses also solutions with a non-vanishing magnetic $SU(2)$ fields.
Remarkably, these emerge as perturbations of the Abelian configurations at some finite temperature
depending on the values of the electric charges, and can be viewed as $p-$wave superfluids.
Moreover, by using the relations in \cite{Lu:1999bw}, \cite{Gauntlett:2007sm}, 
one can uplift these configurations to ten dimensional type IIB supergravity
and $D=11$ supergravity. This provides an explicit stringy construction of holographic superfluids.

Our study should be viewed only as a preliminary investigation
of the  simplest non-Abelian solutions of the ${\cal{N}}=4^+$ model featuring superfluid properties.
Various properties of these black branes remain to be investigated.
For example, it would be interesting to compute the conductivity of the solutions
or to explore the connection with the unbalanced mixtures discussed recently in \cite{Erdmenger:2011hp}.

Moreover, we expect the ${\cal{N}}=4^+$ model to possess a variety of other electrically charged black brane 
solutions\footnote{A class of instabilities of the electrically charged solutions of 
the Romans' model  leading to
holographic helical superconductors have been considered 
in the recent work \cite{Donos:2011ff}.}.
They would be found for a different (typically  more complicated) matter field ansatz than (\ref{SU2-an}).
In this context, it is interesting to study solutions in which the Chern-Simons term 
enters the dynamics\footnote{As seen in the case of the pure EYM-$\Lambda$ model \cite{Brihaye:2009cc}, 
the properties of the spherically symmetric
non-Abelian solutions are very different once one switches on a Chern-Simons term in the action.}.
To this end, we have considered non-Abelian black branes possesing a purely magnetic $SU(2)$ field, with
\begin{eqnarray}
A^{(I)}=w(r)(\delta^{I1}dx+\delta^{I2}dy+\delta^{I3}dz),
\end{eqnarray} 
and an electric $U(1)$ field, $B=B_t(r) dt$. This leads to an isotropic energy-momentum tensor, 
$T_x^x=T_y^y=T_z^z$, in which case  
 a suitable metric Ansatz is given by (\ref{metric-an}) with $f(r)=1$.
 Then (\ref{Einstein-eqs})-(\ref{Tij}) yield five equations of motion
for $m$, $\sigma$, $w$,  $B_t$ and $\phi$ which were solved numerically.
Our results show that the properties of
these solutions  differ substantially from those found in
 the anisotropic case, discussed in Section 4.
First, when treating $w(r)$ as a perturbation around the
Abelian solution (which is (\ref{metricU1xU1-1})-(\ref{metricU1xU1-5}) with $Q=0$), 
the  linearized equation can be solved in closed form.
However, the solution looks very similar to (\ref{pert1}) (with a $\log(r-r_H)$ term),
with the result that no normalizable zero mode is found.
Also, different from the case of anisotropic non-Abelian black branes, 
we could not find non-perturbative solutions with $w(r)\to 0$
as $r\to \infty$. As a result, the mass of the solutions computed according to the counterterm prescription 
given in Section 2, diverges\footnote{Some properties of these  solutions were 
discussed in a more general context in \cite{Radu:2006va}.}.
We therefore conclude that these isotropic black brane non-Abelian solutions cannot be interpreted as holographic superfluids.

However, the situation is likely to be different for a more general case featuring 
an anisotropic $SU(2)$ field and a purely magnetic $U(1)$ field (thus beyond the simple Ansatz (\ref{SU2-an})).
Superconducting black brane solutions of this type have been studied recently
in a truncation of ${\cal{N}}=4^+$ Romans' model with a vanishing dilaton and an arbitrary Chern-Simons
coupling constant \cite{Zayas:2011dw}.

We hope to return elsewhere with a systematic study of these aspects.
\\
\\
%%%%%%%%%%%%%%%%%%%%%%%%%%%%%%%%%%%%%%%%%%%%%%%%%%%%%%%%%%%%%%%%%%%%%%%%%
%%%%%%%%%%%%%%%%%%%%%%%%%%%%%%%%%%%%%%%%%%%%%%%%%%%%%%%%%%%%%%%%%%%%%%%%%
{\bf\large Acknowledgements} 
\\
   This work is carried out in the framework of Science Foundation Ireland (SFI) project
RFP07-330PHY.  ~~
E.R.  gratefully acknowledge support by the DFG,
in particular, also within the DFG Research
Training Group 1620 ''Models of Gravity''. 

%%%%%%%%%%%%%%%%%%%%%%%%%%%%%%%%%%%%%%%%%%%%%%%%%%%%%%%%%%%%%%%%%%%%%%%%%%
 
\end{document}